\documentclass{PoS}
\usepackage{braket}
\usepackage{amsmath,amssymb,mathrsfs,textcomp,amscd,bbm} 

\usepackage{array}

\usepackage{graphicx} 
\DeclareGraphicsExtensions{.pdf,.png,.jpg,.mps,.ps,}
\usepackage{subfigure}

\newcommand{\be}{\begin{equation}}
\newcommand{\ee}{\end{equation}}
\newcommand{\bea}{\begin{eqnarray}}
\newcommand{\eea}{\end{eqnarray}}
\newcommand{\bi}{\begin{itemize}}
\newcommand{\ei}{\end{itemize}}

\newcommand{\MeV}{\,\mathrm{MeV}}

\def\mev{{\rm MeV}}
\def\gev{{\rm GeV}}
\def\tev{{\rm TeV}}

\def\ev{\mathrm{e\kern-0.1em V}}
\def\kev{\mathrm{ke\kern-0.1em V}}
\def\mev{\mathrm{Me\kern-0.1em V}}
\def\gev{\mathrm{Ge\kern-0.1em V}}
\def\tev{\mathrm{Te\kern-0.1em V}}

\title{Masses and decay constants of \boldmath$B^{(*)}_c$ mesons with \boldmath$N_f=2+1+1$ twisted mass fermions}

\ShortTitle{Masses and decay constants of $B^{(*)}_c$ mesons with $N_f=2+1+1$ twisted mass fermions}

\author{Damir Becirevic\\
        Laboratoire de Physique Th\'eorique, CNRS, Univ. Paris-Sud, Univ. Paris-Saclay, Orsay, France\\
        E-mail: \email{damir.becirevic@th.u-psud.fr}}

\author{Aurora Melis\\
        Departamento de F\`isica Te\`orica, Universitat de Val\`encia \& IFIC, Universitat de Val\`encia-CSIC\\
        E-mail: \email{me.aurora.16@gmail.com}}

\author{Lorenzo Riggio\\
        Istituto Nazionale di Fisica Nucleare, Sezione di Roma Tre, Rome, Italy\\
        E-mail: \email{lorenzo.riggio@gmail.com}}

\author{\speaker{Giorgio Salerno}\\
        Universit\'a Roma Tre \& Istituto Nazionale di Fisica Nucleare, Sezione di Roma Tre, Rome, Italy\\
        E-mail: \email{giorgio.salerno.phys@gmail.com}}

\author{Silvano Simula\\
        Istituto Nazionale di Fisica Nucleare, Sezione di Roma Tre, Rome, Italy\\
        E-mail: \email{simula@roma3.infn.it}}

\author{for the ETM Collaboration}

\abstract{We present a preliminary lattice determination of the masses and decay constants of the pseudoscalar and vector mesons $B_c$ and $B_c^*$. Our analysis is based on the gauge configurations produced by the European Twisted Mass Collaboration with $N_f = 2 + 1 + 1$ flavors of dynamical quarks. We simulated at three different values of the lattice spacing and with pion masses as small as 210 MeV. Heavy-quark masses are simulated directly on the lattice up to $\sim 3$ times the physical charm mass. The physical b-quark mass is reached using the ETMC ratio method. Our preliminary results are: $M_{B_c} = 6341\,(60)$ MeV, $f_{B_c} = 396\,(12)$ MeV, $M_{B_c^*} / M_{B_c} = 1.0037\,(39)$ and $f_{B_c^*} /  f_{B_c} = 0.987\,(7)$.}

\FullConference{The 36th Annual International Symposium on Lattice Field Theory - LATTICE2018\\
		22-28 July, 2018\\
		Michigan State University, East Lansing, Michigan, USA.}

\begin{document}

\section{Introduction and simulation details}
\label{intro}

The Standard Model (SM) of particle physics is very powerful in predicting most of the observed phenomena, but it does not provide any explanation for the origin of flavor: the theory parametrizes the  hierarchy of quark masses and CKM mixing angles \cite{Cabibbo:1963yz,Kobayashi:1973fv} through free parameters (6 masses, 3 angles and 1 complex phase).
While the gauge sector is constrained by the $SU(3)_C \otimes SU(2)_L \otimes U(1)_Y$ symmetry, the flavor sector is particularly sensitive to possible extensions of the SM.
The only way to constrain the free parameters of the SM and to investigate New Physics (NP) effects is to combine experimental inputs with theoretical predictions based on first principles.
In this respect a precise determination of the decay constants of the pseudoscalar and vector mesons $B_c$ and $B_c^*$ using QCD simulations on the lattice is crucial. On one hand
$f_{B_c}$ is a very important SM input to constrain possible NP effects implied by the $R(D^{(*)})$ anomalies in the leptonic $B_c \to \tau \nu_{\tau}$ decay and to determine the CKM matrix element $V_{cb}$ with very high accuracy.  
On the other hand, $f_{B_c^*}$ is involved in the description of semileptonic form factors, within the nearest resonance model, and non-leptonic decays through the factorization approximation, so a precise knowledge of $f_{B_c^*}$ is very important today. However, since $f_{B_c^*}$ is not directly measurable in the experiments, its lattice determination is needed to gain access to this quantity.

In this contribution we present a preliminary lattice determination of the masses and decay constants of the pseudoscalar and vector mesons $B_c$ and $B^{*}_c$ using the gauge configurations generated by the European Twisted Mass Collaboration (ETMC) with $N_f = 2 + 1 + 1$ dynamical quarks, which include in the sea, besides two light mass-degenerate quarks, also the strange and the charm quarks \cite{Baron:2010bv,Baron:2011sf}.
The QCD simulations used in this work are the same adopted in Ref.~\cite{Lubicz:2017asp}, where the reader is referred to for details. 

They have been carried out at three different values of the inverse bare lattice coupling $\beta$, at different lattice volumes (with spacial sizes varying between $\simeq 2$ and $\simeq 3$ fm) and for pion masses ranging from $\simeq 210$ to $ \simeq 450$ MeV~\cite{Carrasco:2014cwa}.

We have simulated the heavy-quark mass $m_h$  in the range $m_c < m_h < 3m_c$ (see Table~1 of Ref.~\cite{Lubicz:2017asp}), where $m_c$ is the physical mass of the charm quark.
Three values of the charm mass are used to interpolate to the physical charm quark mass $m_c$, while the physical $b$-quark point $m_h = m_b$, determined in Ref.~\cite{Bussone:2016iua}, is reached by using the \emph{ratio} method of Ref.~\cite{Blossier:2009hg}: an appropriate set of ratios of the masses and decay constants, having a precisely known static limit, is constructed using nearby values of the heavy quark mass. In this way it is possible to control the behavior of the ratios as a function of $m_h$ up to the static limit. The physical $b$-quark point is then reached by performing an interpolation of such ratios at the physical $b$-quark mass $m_b$.
The great advantages of the ratio method are: ~ i) to perform B-physics applying the same relativistic action used for light and charm quarks, and ~ ii) the drastic reduction of both the discretisation errors and the uncertainties related to the perturbative matching between QCD and the heavy-quark effective theory (HQET).

\section{Decay constants and masses on the lattice}
\label{Dc_and_mass_on_lat}

Let us define unphysical heavy-heavy pseudoscalar $H_c$ and a vector $H^*_c$ mesons composed by a valence heavy quark $h$, with a mass $m_h \ge m_c$, and a charm quark.
Since we employ a setup with maximally twisted fermions, the vector, axial and pseudoscalar operators renormalize multiplicatively~\cite{Frezzotti:2003ni}, i.e.~$\widehat{V}_\mu =  {\cal{Z}}_A \cdot V_\mu$ , $\widehat{A}_\mu =  {\cal{Z}}_V \cdot A_\mu$ and $\widehat{P}=  {\cal{Z}}_P\cdot P$, where $V_\mu = \bar{h} \gamma_\mu c$, $A_\mu = \bar{h} \gamma_\mu \gamma_5 c$ and $P = \bar{h} \gamma_5 c$ are the local bare vector, axial and pseudoscalar operators\footnote{The Wilson r-parameters for the heavy and the charm quarks are always chosen to be opposite, i.e.~$r_h = - r_c$.} with ${\cal Z}_V$, ${\cal Z}_A$ and ${\cal Z}_P$ being the corresponding renormalization constants.
The $H_c$ and $H^*_c$ decay constants and masses are related to the matrix elements of $\widehat{V}_\mu$, $\widehat{A}_\mu$ and $\widehat{P}$ as:
\begin{eqnarray}
\label{eq:vector}
\braket{0|\,\widehat{V}_\mu\,|H^*_c(\vec{p},\lambda)} &\!\!\!=\!\!\!& f_{H^*_c} M_{H^*_c} \epsilon^{\lambda}_\mu ~ , \\[2mm]
\label{eq:scalar}
(m_h+m_c)\braket{0|\,\widehat{P}\,|H_c(\vec{p})} & \!\!\!=\!\!\! & p^{H_c}_\mu \braket{0|\,\widehat{A}_\mu\,|H_c(\vec{p})} = f_{H_c} M_{H_c}^2 ~ ,
\end{eqnarray}
where $\epsilon^\lambda$ is the vector meson polarization 4-vector. In Eq.~(\ref{eq:scalar}) we used the axial Ward-Takahashi identity, which is fulfilled also on the lattice in our maximally twisted mass Wilson formulation.
The vector and pseudoscalar matrix elements can be extracted by studying two-point correlation functions
at large time distances, viz.
\begin{eqnarray}
\label{eq:CP} 
C_P(t) & = &\braket{\sum_{\vec{x}}  \widehat{P}(\vec{x},t)\widehat{P}^\dagger(\vec{0},0)} \underset{t\gg a }{\longrightarrow}  
\frac{|\!\braket{0|\widehat{P}(0)|H_c(\vec{0})}\!|^2}{2 M_{H_c}} \left[ e^{-M_{H_c^*} t} + e^{-M_{H_c^*} (T - t)} \right] ~ \\
\label{eq:CV}
C_V(t) & = & \frac{1}{3}\braket{\sum_{i,\vec{x}}  \widehat{V}_i(\vec{x},t)\widehat{V}_i^\dagger(\vec{0},0)} \underset{t\gg a }{\longrightarrow} 
\sum_{i,\lambda}\frac{|\!\braket{0|\widehat{V}_i(0)|H_c^*(\vec{0},\lambda)}\!|^2}{6M_{H_c^*}} \left[ e^{-M_{H_c^*} t} + e^{-M_{H_c^*} (T - t)} \right] ~ ,
\end{eqnarray}
where $T$ is the lattice temporal size.
To improve the determination of $M_{H^{(*)}_c}$ and $f_{H^{(*)}_c}$ we analyzed the whole set of correlators obtained applying local ($L$) and smeared ($S$) interpolators at both the source and the sink, namely $C_{P,V}^{LL}$, $C_{P,V}^{SL}$, $C_{P,V}^{LS}$ and $C_{P,V}^{SS}$.

\begin{figure}[htb!]
\centering
\includegraphics[scale=0.5]{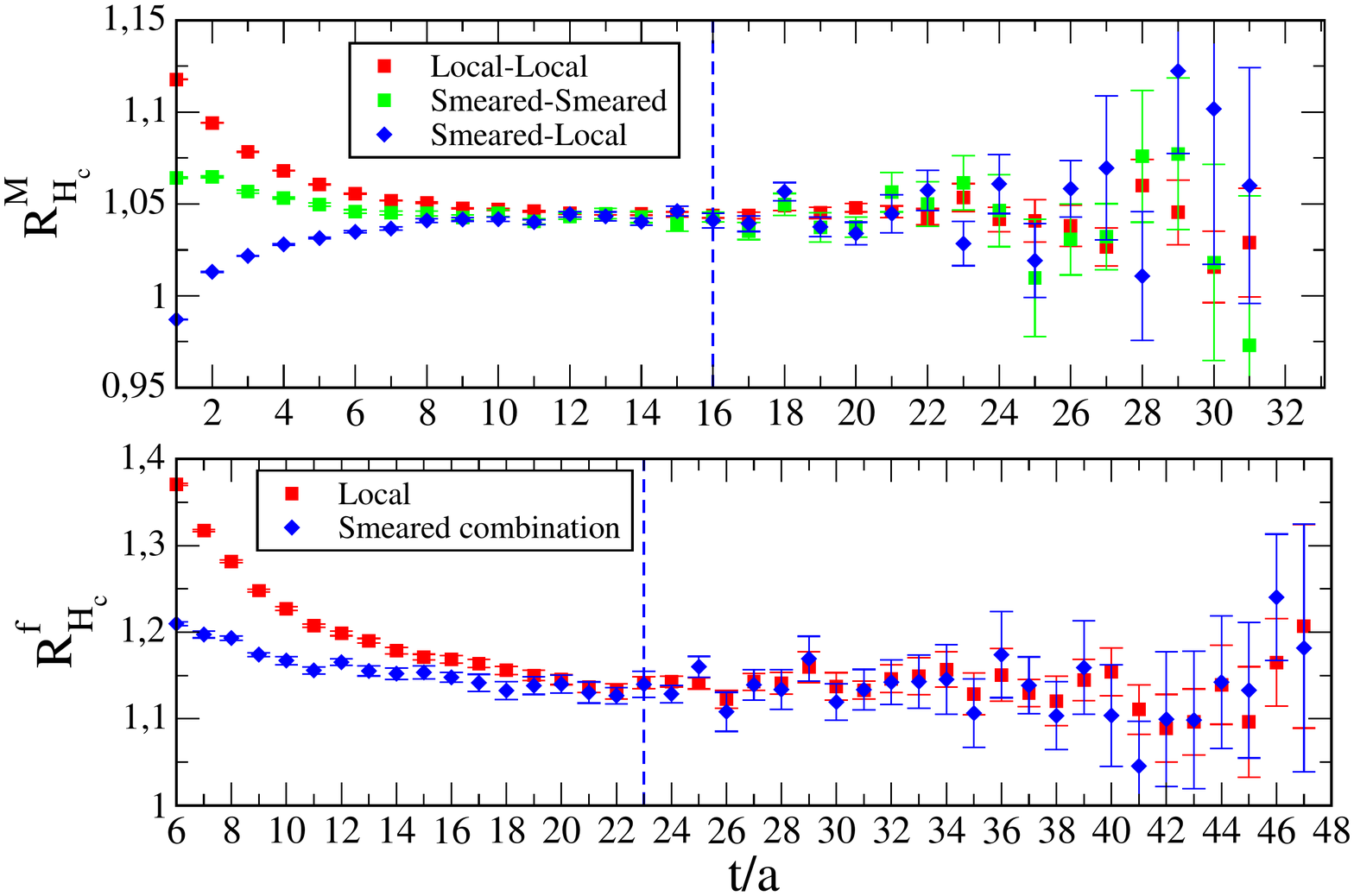}
\vspace{-2cm}
\caption{\it \footnotesize Upper panel: time dependence of the effective mass ratio $R_{\rm eff}$ corresponding to $LL$, $SL$ and $SS$ interpolators in the case of the ensemble A40.32 and for $ m_h \simeq 2.8$ GeV. Lower panel: $R_f^{LL}$ and $R_f^{SL}$ as a function of the time distance for the ensemble A20.48 and $m_h \simeq 1.9$ GeV (see Ref.~\cite{Lubicz:2017asp}).}
\label{fig:Reff_Rf_Plateaux}
\end{figure}

In the the static limit HQET predicts that the vector and pseudoscalar mesons $H_c^*$ and $H_c$, which differ only for their spin configuration, belong to a doublet of spin-flavor symmetry and therefore they have the same mass and decay constant in that limit. 
This means that the ratios $M_{H^*_c}/M_{H_c}$ and $f_{H^*_c}/f_{H_c}$ are equal to one in the static limit.
We exploit this property of HQET by considering the following ratios
\begin{eqnarray}
\label{eq:ratios}
R^M_{H_c}(m_h) =M_{H_c^*} / M_{H_c} \qquad \mbox{and} \qquad R^f_{H_c}(m_h) = f_{H_c^*} / f_{H_c} ~ , 
\end{eqnarray}
where we stress discretization effects are significantly reduced with respect to the case of the individual masses and decay constants. The value of $R^M_{H_c}$ is obtained from the large time distance behavior of the ratio of the pseudoscalar and vector effective masses, namely
\begin{equation}
\label{eq:Reff}
R_{\rm eff}(t) \equiv \frac{M_{\rm eff}^V(t)}{M_{\rm eff}^P(t)} \underset{t \gg a }{\longrightarrow} R^M_{H_c}(m_h, a^2)\,\quad\mbox{with}\quad M_{\rm eff}^{P,V}(t) = \log\left[ \frac{C_{P,V}(t)}{C_{P,V}(t+1)} \right] \underset{t\gg a }{\longrightarrow} M_{H_c^{(*)}}\,.
\end{equation}
To extract $R_{\rm eff}(t)$ we performed a simultaneous fit of $LL$, $SL$, $LS$ and $SS$ correlators, while the decay constant ratio $R^f_{H_c}$ has been determined in two different ways by considering only $LL$ correlators or a combination of $SL$ and $SS$ correlators:
\begin{eqnarray}
\label{eq:Rf_LL}
R^{LL}_{f}(t) &\!\!\!\equiv\!\!\!& \frac{\sinh\left(aM_{H_c}\right)}{am_h + am_c} 
\sqrt{ \frac{C_V^{LL}(t)}{C_P^{LL}(t)} \frac{M_{H_c}}{M_{H_c^*}}  \frac{e^{-M_{H_c}t} + e^{-M_{H_c}(T-t)} }{ e^{-M_{H_c^*}t} + e^{-M_{H_c^*}(T-t)} } } \underset{t\gg a }{\longrightarrow} R^{f}_{H_c}(m_h, a^2) \,,\\
\label{eq:Rf_SL}
R^{SL}_{f}(t) &\!\!\!\equiv\!\!\!& \frac{\sinh\left(aM_{H_c}\right)}{am_h + am_c} \frac{C_V^{SL}(t)}{C_P^{SL}(t)} \sqrt{  \frac{C_P^{SS}(t)}{C_V^{SS}(t)} \frac{M_{H_c}}{M_{H_c^*}} \frac{e^{-M_{H_c}t} + e^{-M_{H_c}(T-t)} }{ e^{-M_{H_c^*}t} + e^{-M_{H_c^*}(T-t)} } } \underset{t\gg a }{\longrightarrow} R^{f}_{H_c}(m_h, a^2) \,.
\end{eqnarray}
The quality of the plateaux for the ratios $R_{\rm eff}$, $R^{LL}_{f}$ and $R^{SL}_{f}$ is shown in Fig.~\ref{fig:Reff_Rf_Plateaux}, while the time intervals used to isolate safely the ground-state are shown in Table~\ref{tab:time_intervals}.

\begin{table}[htb!]
\centering
\small{
		\begin{tabular}{||c|c|c|c|c||}
			\hline
			$\beta$ & $V / a^4$ &$[t_{\rm min},\,t_{\rm max}]_{(LL)}/a$&$[t_{\rm min},\,t_{\rm max}]_{(SL)}/a$&$[t_{\rm min},\,t_{\rm max}]_{(SS)}/a$\\
			\hline
			$1.90$ & $32^{3}\times 64$ &$[19,\,31]$&$[16,\,31]$&$[13,\,31]$ \\
			& $24^{3}\times 48 $ & $[19,\,23]$&$[16,\,23]$&$[13,\,23]$\\
			\hline
			$1.95$ & $32^{3}\times 64$ & $[20,\,31]$&$[17,\,31]$&$[14,\,31]$\\
			& $24^{3}\times 48 $ &$[20,\,23]$&$[17,\,23]$&$[14,\,23]$\\
			\hline
			$2.10$ & $48^{3}\times 96$ &$[27,\,47]$&$[23,\,47]$&$[19,\,47]$\\
			\hline
		\end{tabular}
}
\caption{\it \footnotesize Time intervals adopted in the case of $LL$, $SL$ ($LS$) and $SS$ correlators for the extraction of the ratios $R^M_{H_c}$ and $R^f_{H_c}$ from the large time distance behavior of $R_{\rm eff}(t)$, $R_f^{LL}(t)$ and $R_f^{SL}(t)$.}
\label{tab:time_intervals}
\end{table}

\section{Masses and decay constants of the $B_c^{(*)}$ mesons}
\label{Bc_mass_and_DCs}

\begin{figure}[htb!]
	\centering
	\includegraphics[scale=0.225]{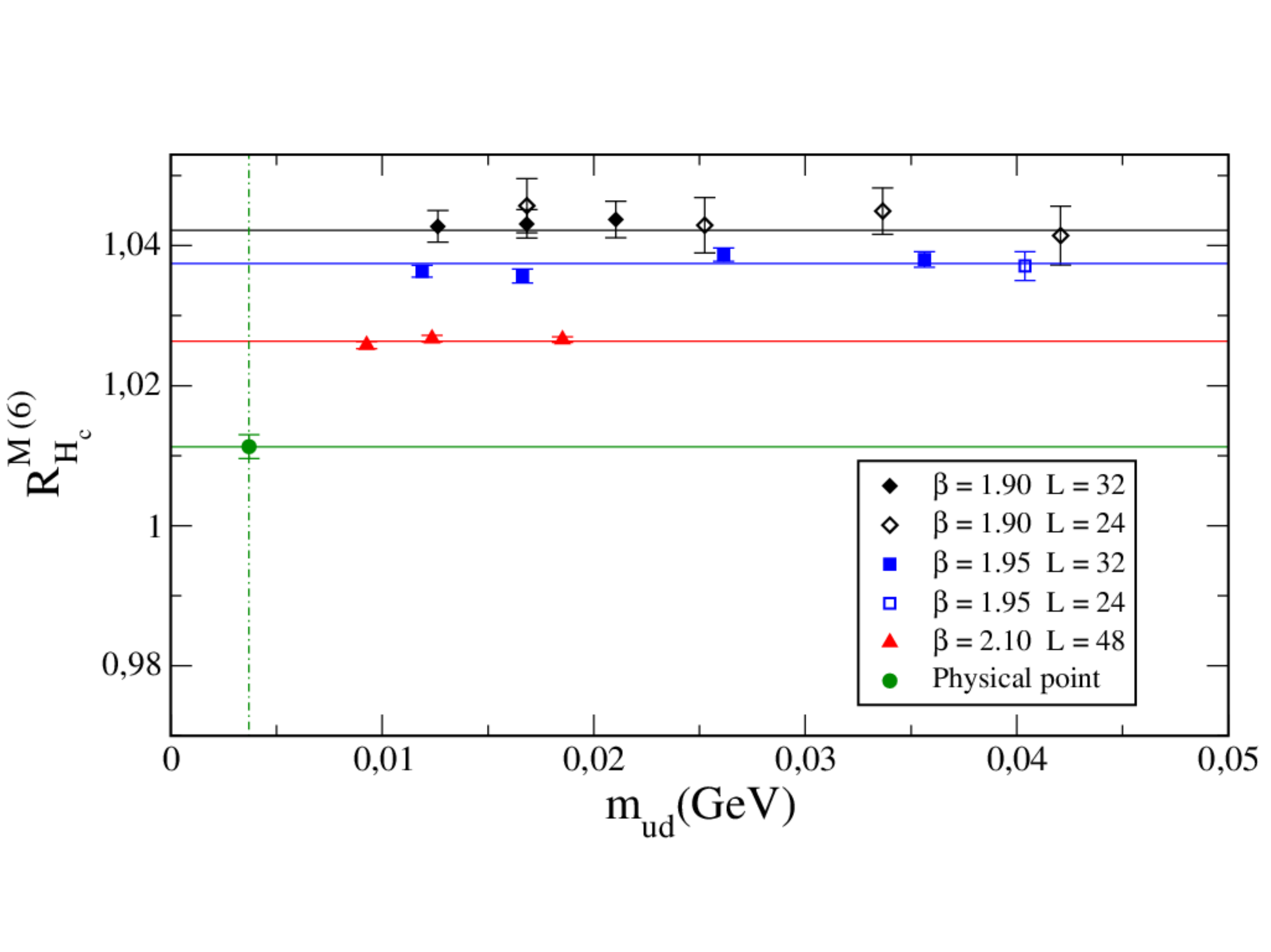}
	\includegraphics[scale=0.225]{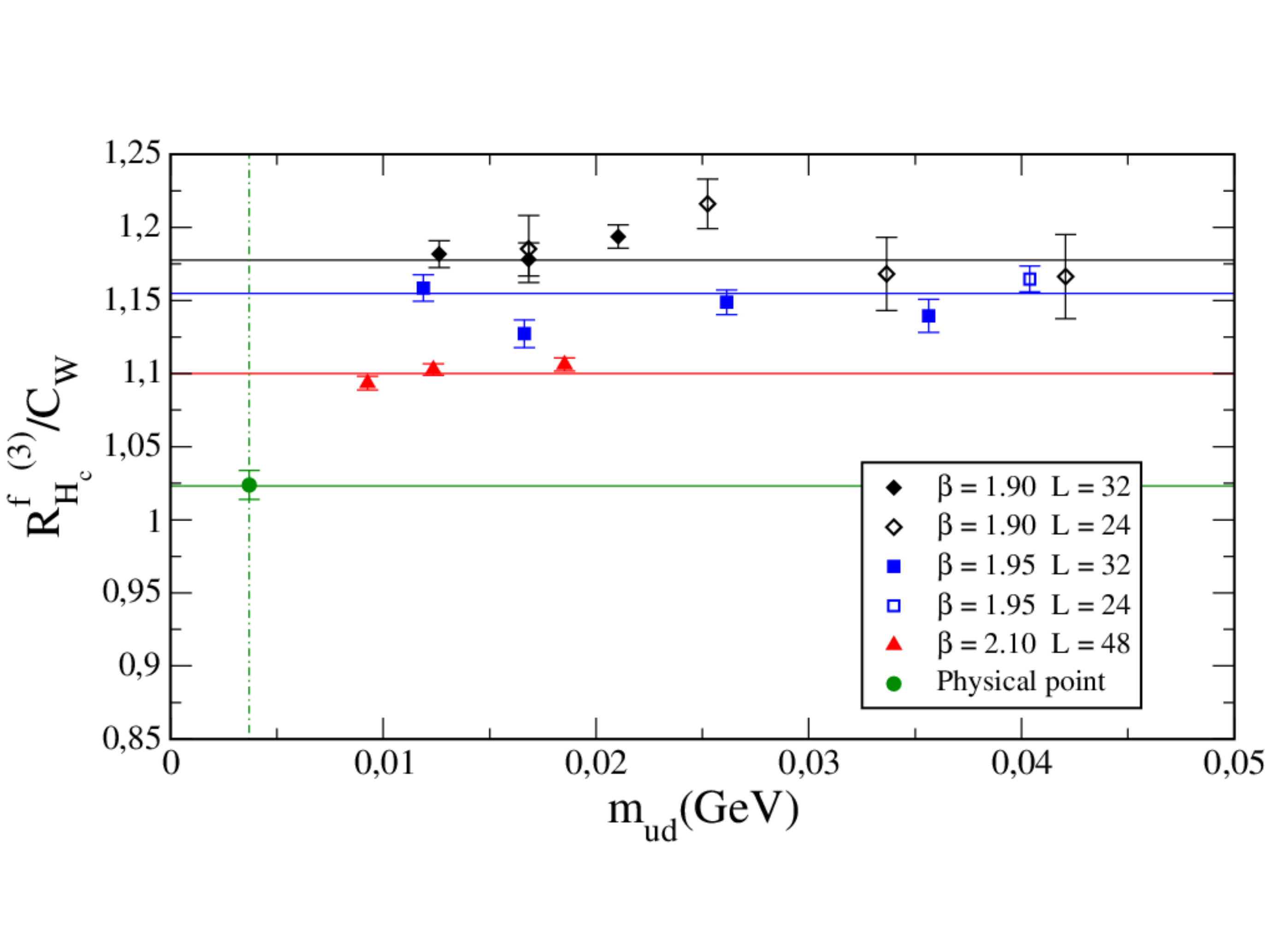}
	\includegraphics[scale=0.225]{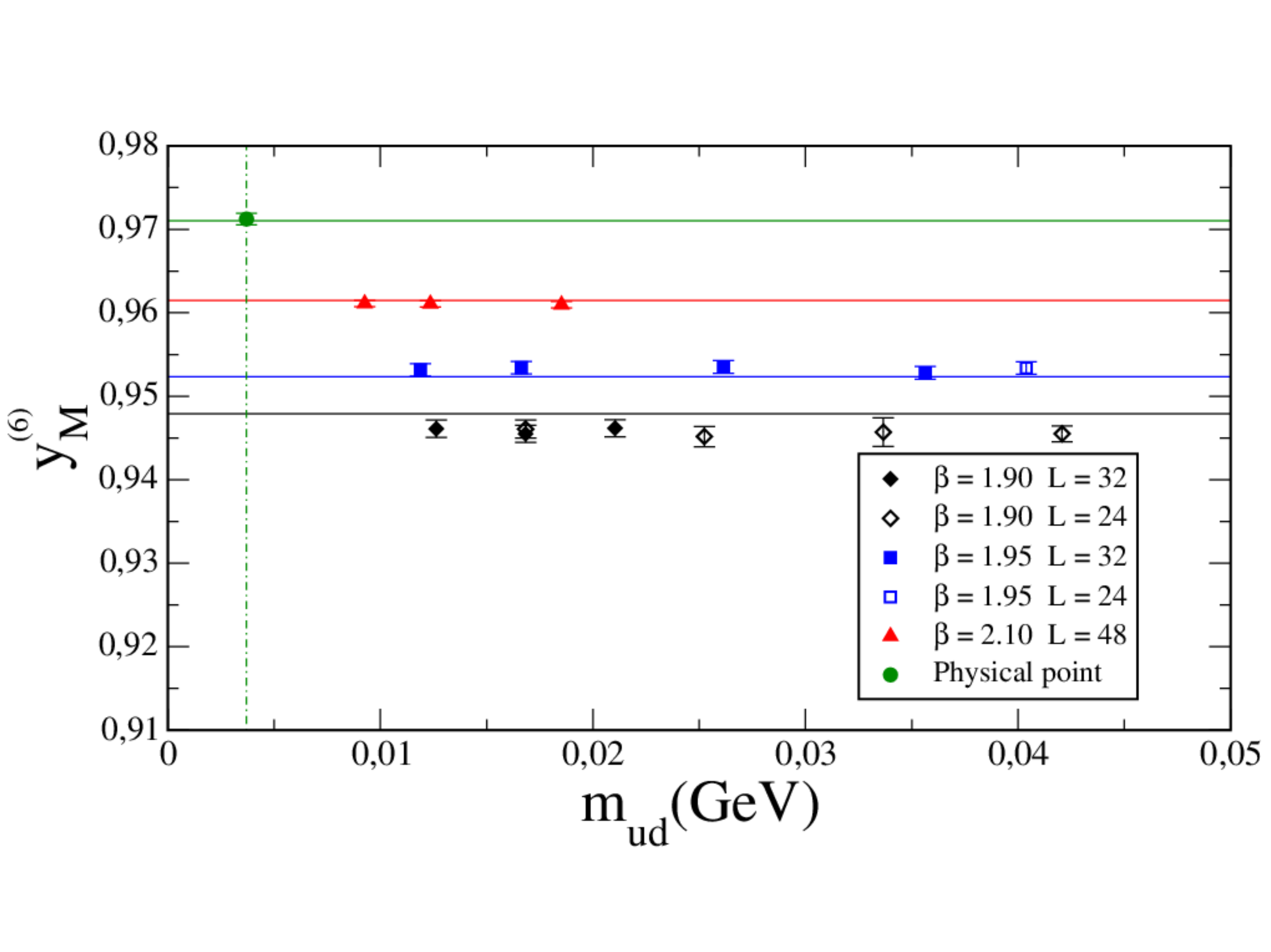}
	\includegraphics[scale=0.225]{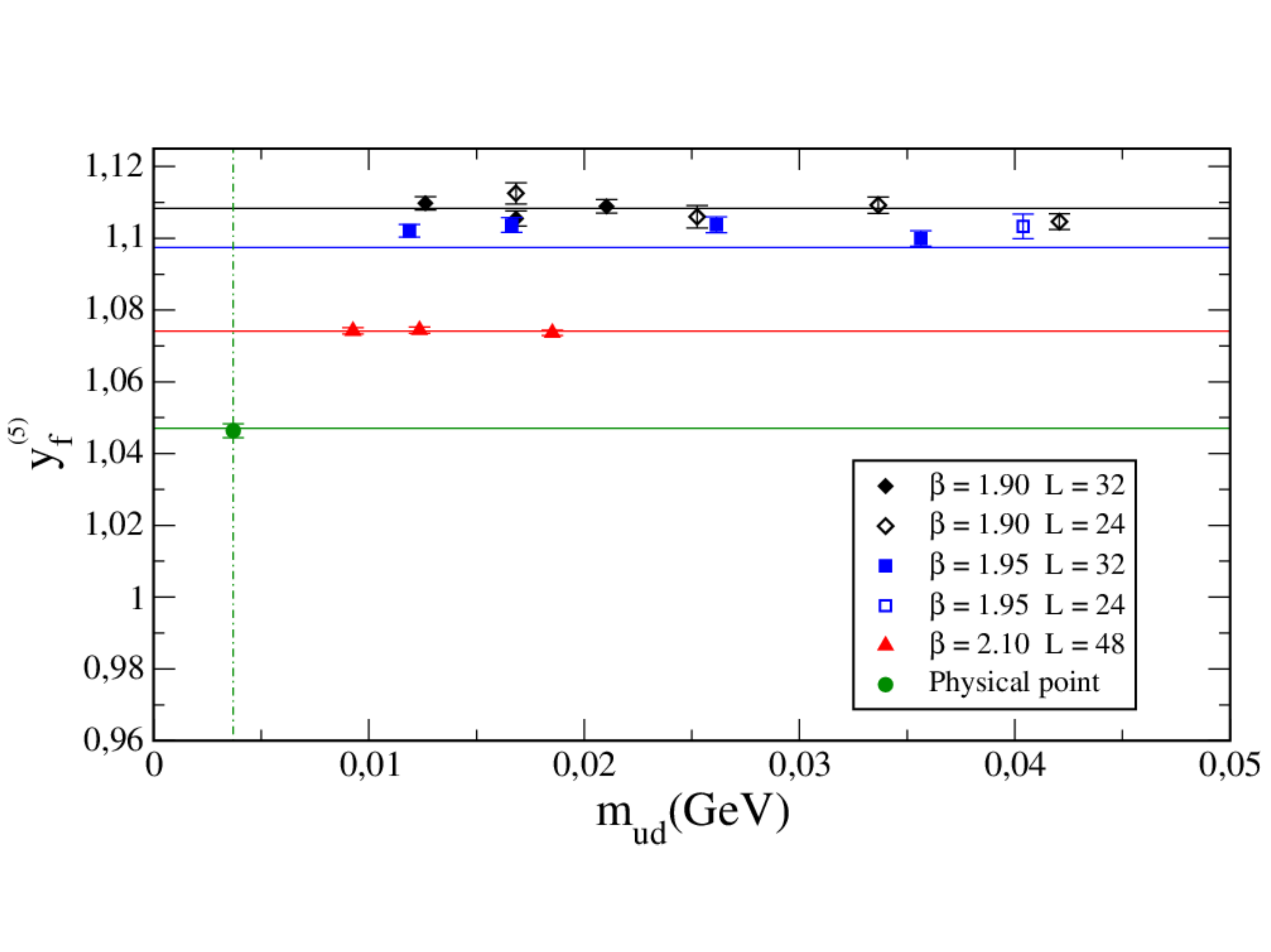}
	\vspace{-0.75cm}
	\caption{\it \footnotesize Combined chiral and continuum extrapolation of the ratios $y_M$ and $y_f$ (see Eqs.(\ref{eq:yM}-\ref{eq:yf})), and of $R^{M}_{H_c}$ and $R^{f}_{H_c}/C_W$ versus the light-quark mass $m_{ud}$. The fit ansatz is described in the text. The green circle represents the result at the physical light-quark mass and in the continuum limit.}
	\label{fig:chir_cont_extrapolation}
\end{figure}

\begin{figure}[htb!]
	\centering
	\includegraphics[scale=0.215]{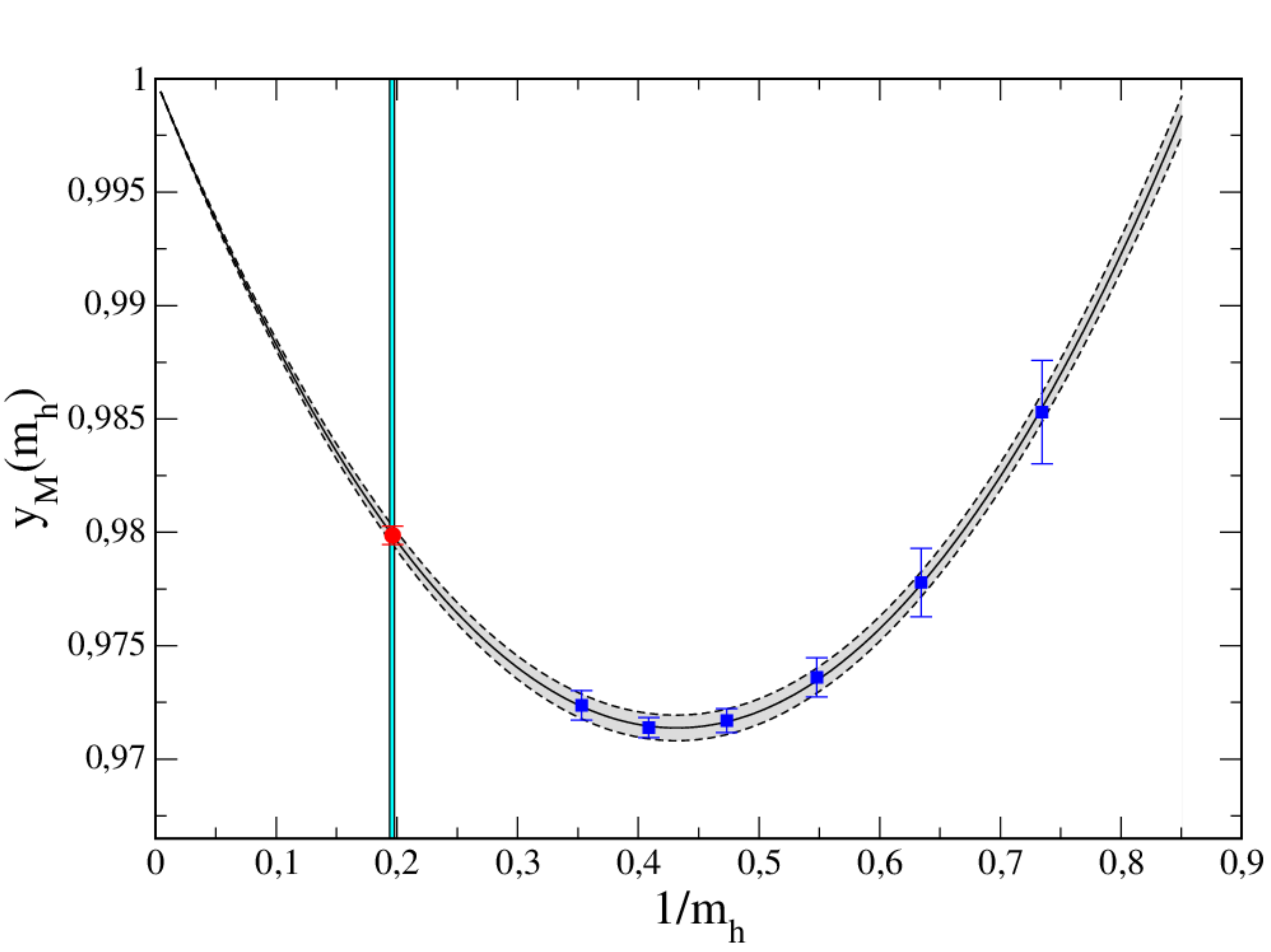}
	\includegraphics[scale=0.215]{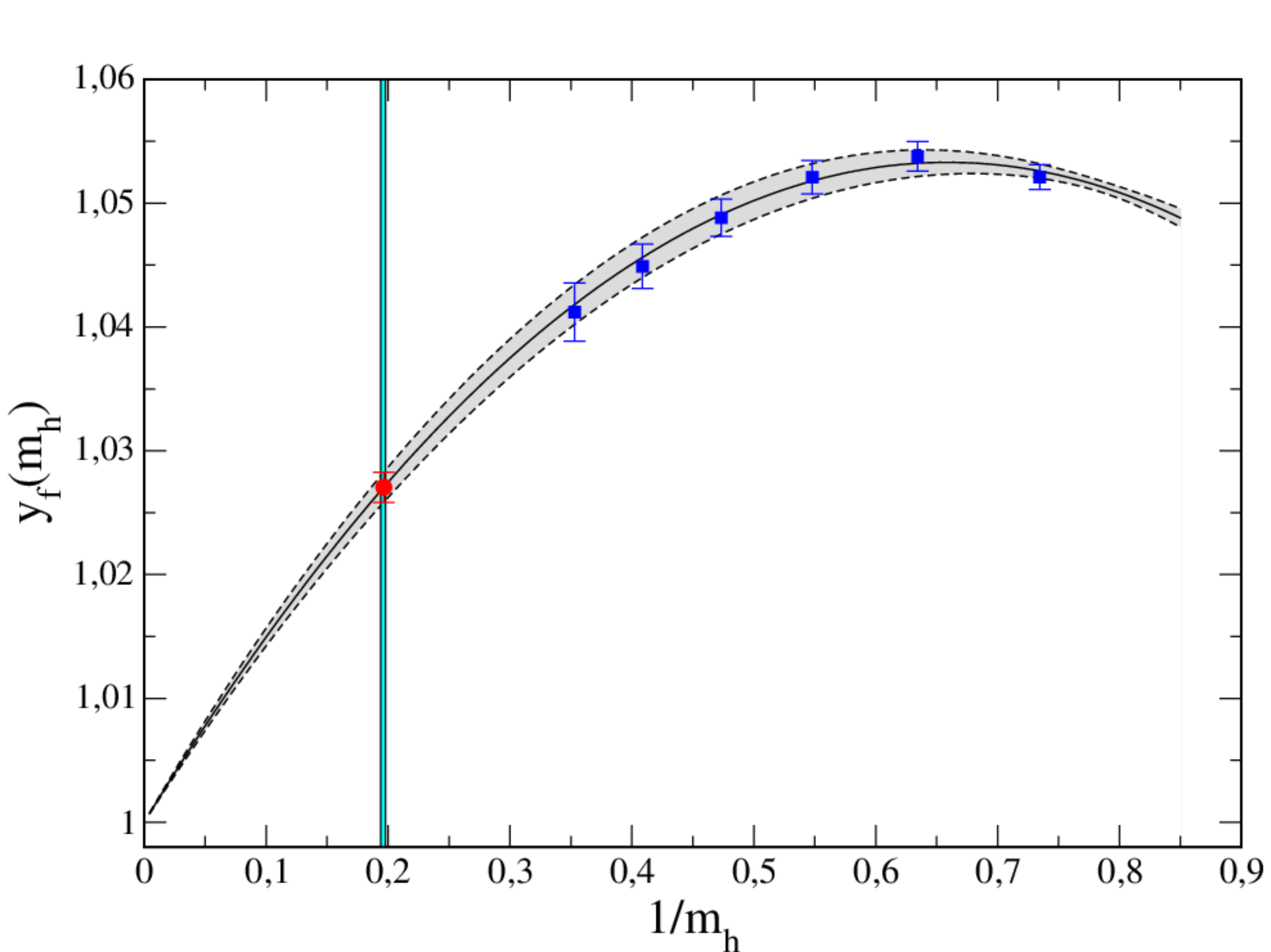}
	\includegraphics[scale=0.215]{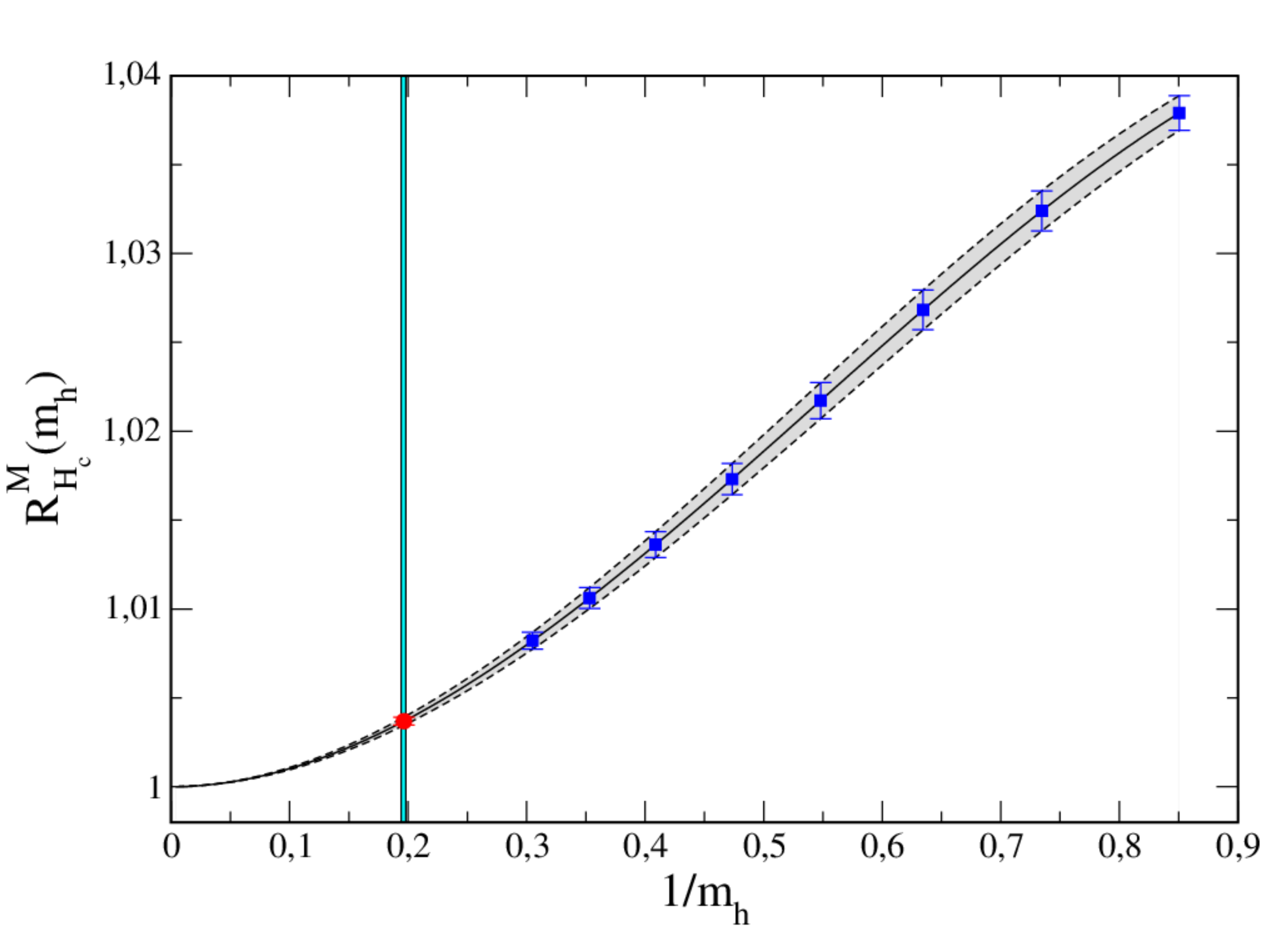}
	\includegraphics[scale=0.215]{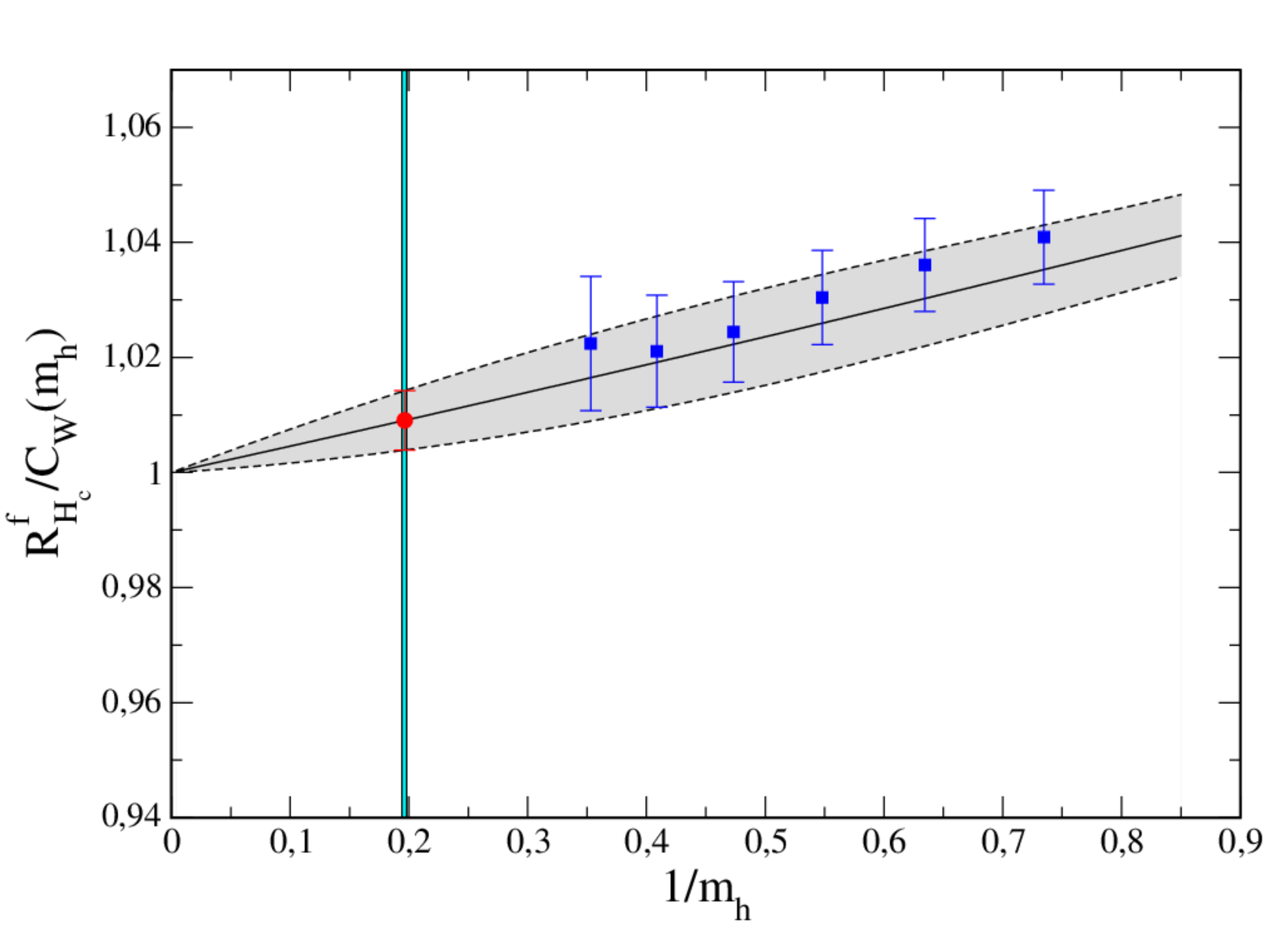}
	\vspace{-0.25cm}
	\caption{\it \footnotesize Dependence of $y_M(m_h)$, $y_f(m_h)$, $R^M_{H_c}(m_h)$ and $R^f_{H_c}(m_h) / C_W$ on the inverse heavy-quark mass $1 / m_h$ and their interpolated values at the physical $b$ quark mass. The interpolations are based on Eqs.~(\ref{eq:mh_dependence_yMf}-\ref{eq:mh_dependence_Rf}).}
	\label{fig:heavy_mass_extrapolation}
\end{figure}

Let us define for each value of the lattice spacing and sea quark mass the following quantities:
\begin{equation}
\label{eq:Qmh_Fmh}
Q(m_h) \equiv M_{H_c}(m_h) / m_h^{pole} \, , \qquad \mathcal{F}(m_h) \equiv f_{H_c}(m_h) \sqrt{m_h^{pole}} / C_A(m_h) \, .
\end{equation}
Using HQET arguments the following set of static limits holds
\begin{eqnarray}
\label{eq:stat_lim}
\underset{m_h \to \infty}{\lim}Q=1 \, , \quad \underset{m_h \to \infty}{\lim}\mathcal{F} = 1 \, \quad
\underset{m_h \to \infty}{\lim}R^{M}_{H_c} = 1 \, , \quad \underset{m_h \to \infty}{\lim}\frac{R^{f}_{H_c}}{C_W} = 1 \, ,
\end{eqnarray}
which will be used to compute the masses and decay constants of the $B_c$ and $B_c^*$ mesons. In Eqs.~(\ref{eq:Qmh_Fmh}-\ref{eq:stat_lim}) $m_h^{pole}$ is the heavy-quark pole mass, while the factors $C_A$ and $C_W$ are the matching coefficients between QCD and HQET, known up to N$^2$LO in perturbation theory~\cite{Beneke:1997jm,Czarnecki:1997vz,Broadhurst:1994se},

We have interpolated our data for $Q$, $\mathcal{F}$, $R^{M}_{H_c}$ and $R^{M}_{H_c}/C_W$ to a sequence of heavy-quark masses that have a common, fixed ratio: $m_h^{(n)}=\lambda m_h^{(n-1)}$ with $n=1, ..., K$, such that $m_h^{(K)}=m_b$.
We fixed the triggering point to the physical charm quark mass, i.e.~$m_h^{(1)}=m_c$. Following Ref.~\cite{Bussone:2016iua} we use $\lambda = 1.160$ and $K = 10$. 
As a second step we constructed at each value of the light-quark mass $m_{ud}$ and lattice spacing the following ratios:
\small
\begin{eqnarray}
\label{eq:yM}
y_M(m_h^{(n)}; m_{ud}, a^2) & = & \frac{Q(m_h^{(n)};\,m_{ud}, a^2)}{Q(m_h^{(n-1)}; m_{ud}, a^2)} = \lambda^{-1}\frac{M_{H_c}(m_h^{(n)}; m_{ud}, a^2)}{M_{H_c}(m_h^{(n-1)}; m_{ud}, a^2)}\frac{\rho(m_h^{(n)}/\lambda,\,\mu)}{\rho(m_h^{(n)},\,\mu)}\,,\\
\label{eq:yf}
y_f(m_h^{(n)}; m_{ud}, a^2) & = & \frac{\mathcal{F}(m_h^{(n)}; m_{ud}, a^2)}{\mathcal{F}(m_h^{(n-1)}; m_{ud}, a^2)} = \lambda^{1/2}\frac{f_{H_c}(m_h^{(n)}; m_{ud}, a^2)}{f_{H_c}(m_h^{(n-1)}; m_{ud}, a^2)}\frac{C_A(m_h^{(n-1)})}{C_A(m_h^{(n)})}\frac{[\rho(m_h^{(n)},\,\mu)]^{1/2}}{[\rho(m_h^{(n-1)},\,\mu)]^{1/2}} ~ , \qquad
\end{eqnarray}
\normalsize
where we have used the relation $m_h^{pole} = m_h \, \rho(m_h, \mu)$ between the pole quark mass and the renormalized quark mass $m_h$ (in the $\overline{\mbox{MS}}$ scheme at the scale $\mu$).
Since the above quantities are double ratios taken at nearby values of the heavy-quark mass $m_h$, the systematic uncertainties due to discretization effects and to the use of the perturbative factor $\rho(m_h, \mu)$ are suppressed even for large values of $m_h$. Thus, we can safely perform the chiral and continuum extrapolation for $y_M$, $y_f$, $R^M_{H_c}$ and $R^f_{H_c} / C_W$ for each value $m_h^{(n)}$. The quality of the scaling behaviour of these ratios is shown in Fig.~\ref{fig:chir_cont_extrapolation}.
Since discretization errors increase unavoidably as the heavy-quark mass gets higher, we have considered the following ansatz for the chiral and continuum extrapolations:
\begin{eqnarray}
\label{eq:chir_cont_yMf}
y_{M,f}(m_h,a^2) & = & y_{M,f}(m_h) \left[ 1 + P_1^{M,f} m_h a^2 + P_2^{M,f} (m_h^{(n)}- m_c) a^4 \right] ~ , \\
\label{eq:chir_cont_RM}
R^M_{H_c}(m_h,a^2) & = & R^M_{H_c}(m_h) \left[ 1+ D_1^M m_h a^2 + D_2^M (m_h^{(n)} - m_c) a^4 \right] ~ , \\
\label{eq:chir_cont_Rf}
R^f_{H_c}(m_h,a^2)/C_W(m_h) & = & R^f_{H_c}(m_h)/C_W(m_h) \left[ 1+ D_1^f m_h a^2 + D_2^f (m_h^{(n)} - m_c) a^4 \right] ~ ,
\end{eqnarray}
where $y_{M,f}(m_h,0)\equiv y_{M,f}(m_h)$, $R^{M(f)}_{H_c}(m_h,0)\equiv R^{M(f)}_{H_c}(m_h)$, while $P_i^{M,f}$ and $D_i^{M(f)}$ ($i=1,2$) are free parameters. 
From the asymptotic behaviors (\ref{eq:stat_lim}) it follows that
\begin{eqnarray}
\label{eq:mh_dependence_yMf}
y_{M,f}(m_h) & = & 1 + A_1^{M,f} / m_h + A_2^{M,f} / m_h^2\,,\\
\label{eq:mh_dependence_RM}
R^M_{H_c}(m_h) & = & 1 + B_1^M / m_h^2 + B_2^M / m_h^3\,,\\
\label{eq:mh_dependence_Rf}
R^f_{H_c}(m_h)/C_W(m_h) & = & 1 + B_1^f / m_h + B_2^f / m_h^2\,,
\end{eqnarray}
where the constraints coming from the static limit are properly incorporated. In this way our data can be interpolated in $1/m_h$ to reach the physical $b$-quark mass as shown in Fig.~\ref{fig:heavy_mass_extrapolation}.
Finally, we can compute $M_{B_c}$ and $f_{B_c}$ through the chain equations:
\begin{eqnarray}
\label{eq:chain_MBc}
M_{\eta_c} y_{M}(m_h^{(1)})y_{M}(m_h^{(2)}) \dots y_{M}(m_h^{(K)}) & = & \lambda^{-K} M_{B_c} ~ \rho(m_h^{(0)},\,\mu) / \rho(m_h^{(K)},\,\mu) ~ , \\
\label{eq:chain_fBc}
f_{\eta_c} y_{f}(m_h^{(1)})y_{f}(m_h^{(2)}) \dots y_{f}(m_h^{(K)}) & = & \lambda^{K/2} f_{B_c} ~ C_A(m_h^{(0)}) \rho(m_h^{(K)},\,\mu) / C_A(m_h^{(K)}) \rho(m_h^{(0)},\,\mu) ~ , ~ \qquad
\end{eqnarray}
where at the physical charm point\footnote{For the $J / \psi$ meson we get the preliminary values $M_{J / \psi} / M_{\eta_c} = 1.0354\,(21)$ and $f_{J / \psi} / f_{\eta_c} = 1.0409\,(80)$.} we obtain $M_{\eta_c} = 2975\,(50)$ MeV and $f_{\eta_c} = 391\,(9)$ MeV.
The values of $M_{B_c^*}$ and $f_{B_c^*}$ can be determined from $M_{B_c^*} = M_{B_c} R^{M}_{H_c}(m_b)$ and $f_{B_c^*} = f_{B_c} R^{f}_{H_c}(m_b)$.

\section{Results and conclusions}
\label{results}

From the analysis of Sec.~\ref{Bc_mass_and_DCs} our preliminary results for the masses and decay constants of the $B_c$ and $B_c^*$ mesons are:
\begin{eqnarray}
\label{eq:MBc_fBc}
M_{B_c} = 6341\,(60)\MeV ~ , && \qquad f_{B_c} = 396\,(12)\MeV\,,\\
\label{eq:ratio_M_f}
M_{B_c^*}/M_{B_c} = 1.0037\,(39) ~ , && \qquad f_{B_c^*}/f_{B_c} = 0.978\,(7)\,.
\end{eqnarray}
Our predictions for $M_{B_c}$ and $M_{B_c^*} / M_{B_c}$ are respectively consistent within $\simeq 1 \sigma$ with the experimental mass from the PDG~\cite{Tanabashi:2018oca} and the recent lattice result $M_{B_c^*} / M_{B_c} = 1.0088\,(16)$ from Ref.~\cite{Mathur:2018epb}. Our value of $f_{B_c^*} / f_{B_c}$ agrees with the result $f_{B_c^*} / f_{B_c} = 0.988\,(27)$ obtained by the HPQCD collaboration~\cite{Colquhoun:2015oha}, while for $f_{B_c}$ there is a $\simeq 2 \sigma$ tension with the HPQCD result $f_{B_c} = 434\,(15)$ MeV. 
Combining Eqs.~(\ref{eq:MBc_fBc}-\ref{eq:ratio_M_f}) we obtain the predictions 
\begin{equation}
\label{eq:MBc_fBc_star}
M_{B_c^*} = 6365\,(65)\MeV ~ , \qquad f_{B_c^*} = 387\,(12)\MeV\,.
\end{equation}

\bibliographystyle{JHEP}
\bibliography{lattice2018.bib}

\end{document}